\documentclass[sigconf]{acmart}
\usepackage{algorithm}
\usepackage{algorithmic}
\AtBeginDocument{%
  }

\setcopyright{acmlicensed}
\acmConference[Conference acronym 'XX]{Make sure to enter the correct
  conference title from your rights confirmation email}{June 03--05,
  2018}{Woodstock, NY}
\acmISBN{978-1-4503-XXXX-X/2018/06}

\begin{document}

\title{MatrixFlow: System-Accelerator co-design \\
for high-performance transformer applications}

\author{Qunyou Liu}
\affiliation{%
  \institution{École Polytechnique Fédérale de Lausanne (EPFL), Embedded Systems Laboratory (ESL)}
  \city{Lausanne}
  \country{Switzerland}}
\email{qunyou.liu@epfl.ch}

\author{Marina Zapater}
\affiliation{%
  \institution{HES-SO University of Applied Sciences and Arts Western Switzerland, Institute of Reconfigurable \& Embedded Digital Systems (REDS)}
  \city{Yverdon-les-Bains}
  \country{Switzerland}}
\email{marina.zapater@heig-vd.ch}

\author{David Atienza}
\affiliation{%
  \institution{École Polytechnique Fédérale de Lausanne (EPFL), Embedded Systems Laboratory (ESL)}
  \city{Lausanne}
  \country{Switzerland}}
\email{david.atienza@epfl.ch}

\renewcommand{\shortauthors}{Trovato et al.}

\begin{abstract}
Transformers are central to advances in artificial intelligence (AI), excelling in fields ranging from computer vision to natural language processing. Despite their success, their large parameter count and computational demands challenge efficient acceleration.
To address these limitations, this paper proposes MatrixFlow, a novel co-designed system-accelerator architecture based on a loosely coupled systolic array including a new software mapping approach for efficient transformer code execution. MatrixFlow is co-optimized via a novel dataflow-based matrix multiplication technique that reduces memory overhead.
These innovations significantly improve data throughput, which is critical for handling the extensive computations required by transformers. We validate our approach through full system simulation using gem5 across various BERT and ViT Transformer models featuring different data types, demonstrating significant application-wide speed-ups. Our method achieves up to a 22x improvement compared to a many-core CPU system, and outperforms the closest state-of-the-art loosely-coupled and tightly-coupled accelerators by over 5x and 8x, respectively.
\end{abstract}



\maketitle

\section{Introduction}
In the realm of high-performance computing, the demand for faster data processing and higher throughput has led to significant advancements in hardware accelerators, including GPUs~\cite{gpu}, Systolic Arrays (SAs)~\cite{b2} and Vector Processors~\cite{vecpro}. These are designed to perform certain types of computations more efficiently than general-purpose CPUs. Among them, SAs have emerged as a crucial architecture for matrix-related computations, prevalent in applications ranging from signal processing to deep learning~\cite{b1}.
A SA is a grid of processors that rhythmically compute and pass data through the system, much like the human circulatory system. This structure is particularly effective for algorithms that can be expressed as matrix operations, allowing for high degrees of parallelism and efficient data movement. 

Transformers have gained significant attention in recent years for their superior performance in various Artificial Intelligence (AI) tasks, including Natural Language Processing (NLP) and computer vision. These models rely heavily on matrix multiplications, making them well suited for implementation on SAs~\cite{b2}. The self-attention mechanisms of the transformers make them especially suited for the high throughput and parallelism offered by SAs. Therefore, integration of transformers with optimized SAs can lead to substantial performance improvements, enabling faster and more efficient processing of large data sets and complex models.

Despite their advantages, traditional implementations of SAs face several limitations that hinder performance. One major challenge is effectively managing the data traffic between memory resources and computational units, when the accelerator is connected in a loosely-coupled way (i.e., directly to the system bus). This often leads to bottlenecks, high latency, and under-utilization of processing capabilities~\cite{b3}. On the other hand, as shown in previous work, tightly-coupling of SAs with CPUs (i.e., as part of the CPU pipeline) introduces overheads from instruction processing and data fetching, which can degrade the overall system performance~\cite{b4}.

To address these limitations, we introduce MatrixFlow, a novel accelerator-system co-optimized architecture with an advanced data structure and algorithm specifically designed for matrix multiplication on systolic arrays (SAs) connected in a loosely-coupled way via standard system bus interconnects. MatrixFlow enhances traditional accelerators by transforming conventional matrix multiplication into a data-flow-based process, significantly improving data handling efficiency and throughput. In this work, we evaluate MatrixFlow across various data formats and system configurations to thoroughly explore its potential performance gains in diverse scenarios. Our contributions are summarized as follows:


\begin{enumerate}
    \item  \textbf{System-Level Co-Optimization of Data Structures and Accelerator Architecture}: A novel data structure and algorithm are designed to optimize data sizes and arrangements that ensure efficient storage and minimal data traffic between memory and computation units. This co-design significantly reduces instruction overhead, enhancing throughput and scalability.

    \item \textbf{Enhancements to Systolic Array Accelerators}: The systolic array accelerator is co-designed with novel data structure and computer system, including PCI Express (PCIe), System Memory Management Unit (SMMU), and Direct Memory Access (DMA). Using the proposed hardware features fully, we optimized data movement through the hardware pipeline. This co-optimization ensures maximal utilization of the hardware's capabilities, leading to significant improvements in data handling efficiency and overall system throughput.
    \item \textbf{Performance Assessment}: Matrix multiplication and transformers are implemented on the accelerator, and their performance is compared against traditional CPU-only setups and other accelerators, demonstrating a performance improvement of up to 698x compared to single-core CPU, and 5x/8x against other loosely- and tightly-coupled accelerators, respectively.
\end{enumerate}

\section{Background and Related Work}

\subsection{Transformers and Systolic Arrays}
The transformer model, introduced by Vaswani et al~\cite{b5}., has revolutionized natural language processing by using a self-attention mechanism to evaluate the importance of words across positions in a sentence~\cite{b5}. Unlike previous models that process inputs sequentially, transformers process the entire sequence simultaneously, enhancing parallelization. 
The main bottleneck of transformers lies in the general matrix multiply (GEMM). Previous works in the state-of-the-art show how more than 99\% of time in a sequential non-accelerated version of the transformer is spent in GEMMs~\cite{b2}.

\begin{figure}[h]
\vspace{-0.4cm}
\centering
\includegraphics[width=1.0\linewidth]{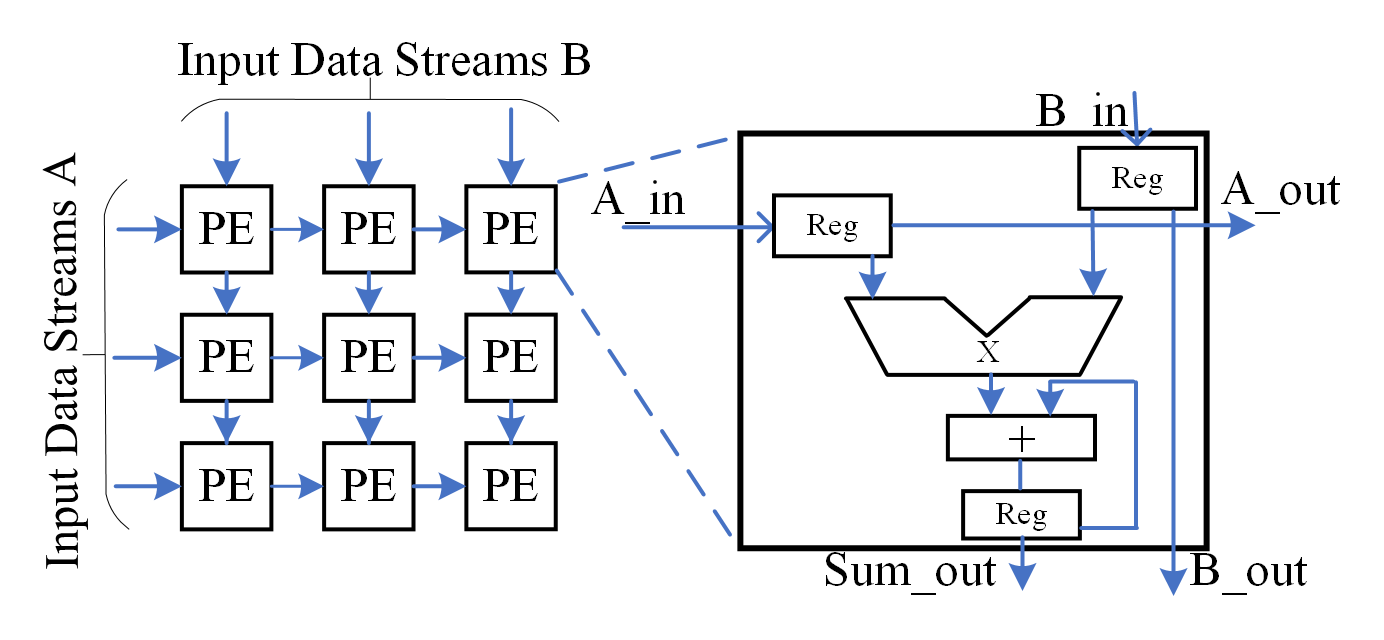}
\vspace{-0.5cm}
\caption{Systolic Array Architecture.}
\label{fig:systolicarray1}
\vspace{-0.5cm}
\end{figure}

Systolic arrays (SAs), widely used in digital signal processing, consist of a grid of processors that handle synchronized parallel data processing. These arrays are highly efficient in AI computation for performing matrix multiplication operations essential for models such as Transformers~\cite{b6}~\cite{b7}. Each processor in a SA not only performs calculations, but also simultaneously passes data to its neighbors, thus facilitating fast data processing and reduced power consumption, as shown in Fig.~\ref{fig:systolicarray1}. 
Data flow through the array horizontally and vertically, and partial sums are accumulated inside each processing element (PE). 
The inherent ability of the SA to perform multiple operations simultaneously makes it an ideal candidate for accelerating tasks in AI algorithms, particularly those involving large matrix multiplications.

\subsection{Advancements and Gaps in AI Hardware Acceleration}
The re-emergence of SA accelerators for AI lies in their efficiency in managing data flow~\cite{b8}. However, they face challenges in scalability and flexibility, which limits their adaptability to various AI tasks. On the computational front, recent work optimizes transformers, with a focus on enhancing the computational speed of the accelerator itself while often neglecting the data flow and memory efficiency of the system~\cite{b1}.

A critical bottleneck in transformer performance is data movement rather than computation itself. Hoefler et al.~\cite{b9} demonstrate that optimizing data movement is essential to enhance the overall performance of large AI models. Their study emphasizes that inefficiencies in data transfer and memory access patterns can severely limit the throughput of transformers, underscoring the need for advanced data management strategies to fully exploit the computational capabilities of modern hardware. However, they focus mainly on the software optimization and target transformer training on GPUs.

To optimize the transformer performance, matrix tiling has emerged as an essential technique for improving the bottlenecks in GEMMs. For example, Miniskar et al. propose a tiling framework that supports heterogeneous memory mapping and uses native APIs for architecture-supported tiled data transfer, which significantly improve performance over conventional methods~\cite{b10}. However, they mainly focus on conventional CPU-centric systems. Another method of optimizing the transformer is a specific accelerator design. Recently, a tightly-coupled accelerator (i.e., directly connected to the CPU pipeline) was proposed to optimize GEMMs in transformers~\cite{b2}. The result shows a tremendous improvement compared to conventional single-core CPU systems. However, implementing the accelerator requires the modification of both the ISA and the execution stage of the CPU pipeline.

Our research aims to fill these gaps by proposing a novel data structure and matrix multiplication method, and loosely-coupled (i.e., connected to the system bus, without modifications to the CPU itself) SA design that incorporates advanced PCIe and DMA capabilities, optimizing both the computational speed and system-wide data management.

\section{System and MatrixFlow}
We co-design both the system and the accelerator to increase the performance of GEMM based on a stardard PCIe interconnect. We propose \textit{MatrixFlow}, an accelerator featuring a novel matrix multiplication method optimized for data-flow accelerators, along with an innovative data structure to improve data streaming efficiency and memory utilization.

\subsection{Accelerator Hardware Design}
\begin{figure}[h]
\vspace{-0.2cm}
  \centering
  \includegraphics[width=0.65\linewidth]{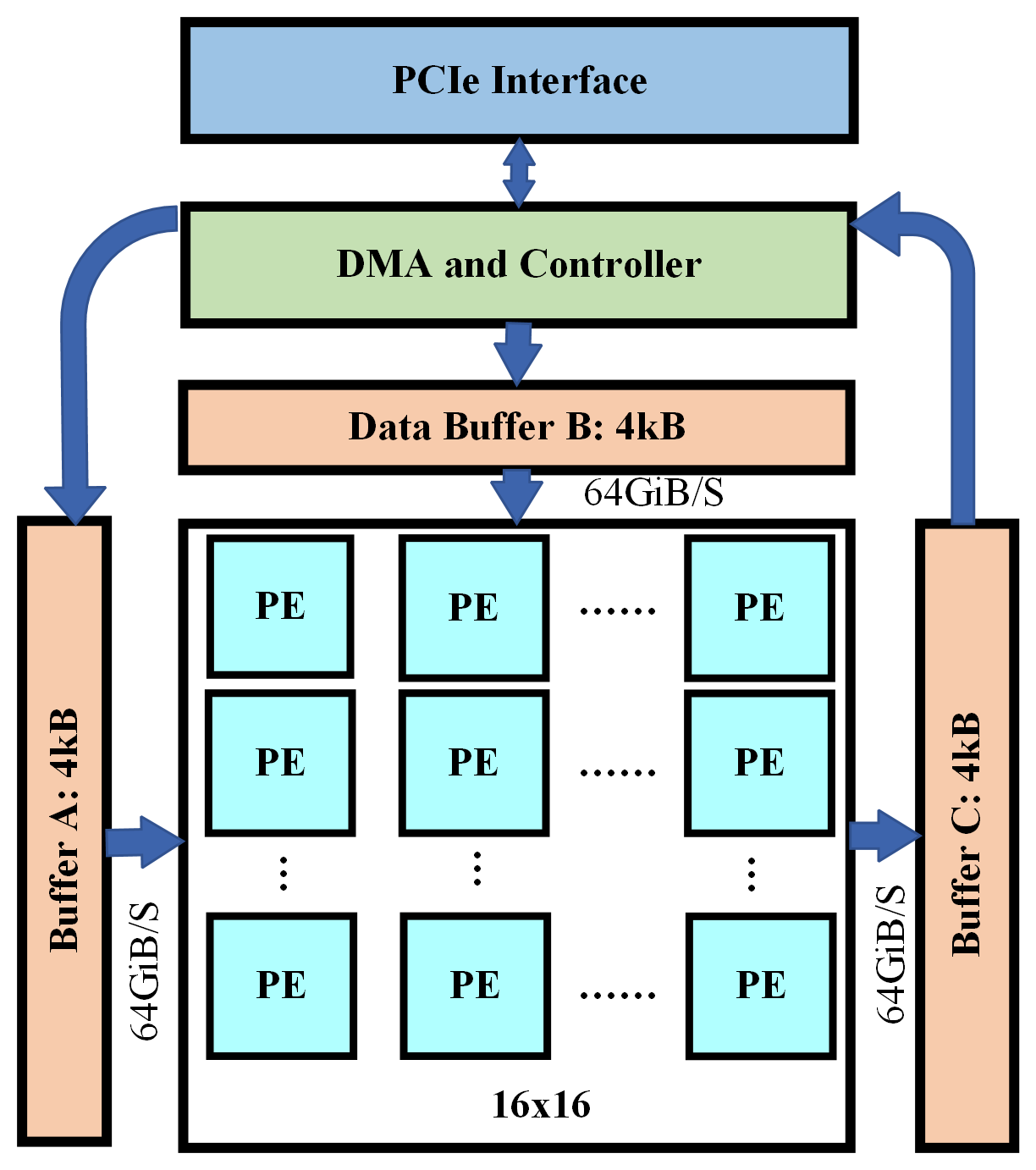}
  \vspace{-0.25cm}
  \caption{Hardware design for systolic array}
  \label{fig:sysmy}
  \vspace{-0.2cm}
\end{figure}

The SA designed for matrix multiplication is depicted in Figure~\ref{fig:sysmy}. Unlike other state-of-the-art designs, we enhance the conventional processing element (PE) structure of an SA (highlighted in blue) by integrating components required to establish a loosely-coupled connection via a PCIe interface (shown in blue and green) within a system utilizing paging (depicted in orange). Compared to other open-source systolic-array-based accelerators, our design incorporates a significantly smaller local buffer. Specifically, we used only three 4kB local buffers, in contrast to the 256kB~\cite{gemmini}, 128kB~\cite{sysacc1}, and 96kB~\cite{sysacc2} buffers used in other works. This reduction is enabled by our co-optimized system-level data structure and algorithm design(see~\ref{sec:acceopt_mm}).
Our design supports multiple data types, including \texttt{INT32}, \texttt{INT16}, \texttt{INT8}, \texttt{FLOAT32}, and \texttt{FLOAT16}. To accommodate the unique requirements of each data type, we implemented different hardware designs, allowing optimized performance and area efficiency. This enables us to analyze and compare the area and performance trade-offs across different precisions. Analyzing the impact of precision and quantization on model accuracy is beyond the scope of this paper. Therefore, we utilize data types that result in quantized transformer models with negligible accuracy drops according to the literature~\cite{b12}.

\textbf{Buffer Configuration:}
\begin{itemize}
    \item \textit{Buffers A and B:} Hold 4KB of data each, corresponding to the page size of a modern computer system, for input matrices A and B, respectively.
    \item \textit{Buffer C:} Collects the output, initiating a DMA transfer to system memory once full.
\end{itemize}

\textbf{Data Flow and Processing:}
\begin{itemize}
    \item \textit{Data Entry and Propagation:} The inputs from matrices A and B enter horizontally and vertically, moving through a 16x16 grid of PE units that multiply and accumulate.
\end{itemize}

\textbf{System Control and Management:}
\begin{itemize}
    \item \textit{DMA Integration:} It manages data transactions between buffers and system memory, optimizing throughput and latency to match the data flow explained in \ref{sec:acceopt_mm}.
    \item \textit{Hardware Interrupts:} Triggers an interrupt upon processing completion to manage subsequent tasks.
\end{itemize}

This design improves matrix multiplication efficiency and optimizes data flow, ensuring high performance and low latency for demanding computing environments.

\subsection{System and Kernel Design}
\begin{figure}[h]
\vspace{-0.4cm}
  \centering
  \includegraphics[width=0.75\linewidth]{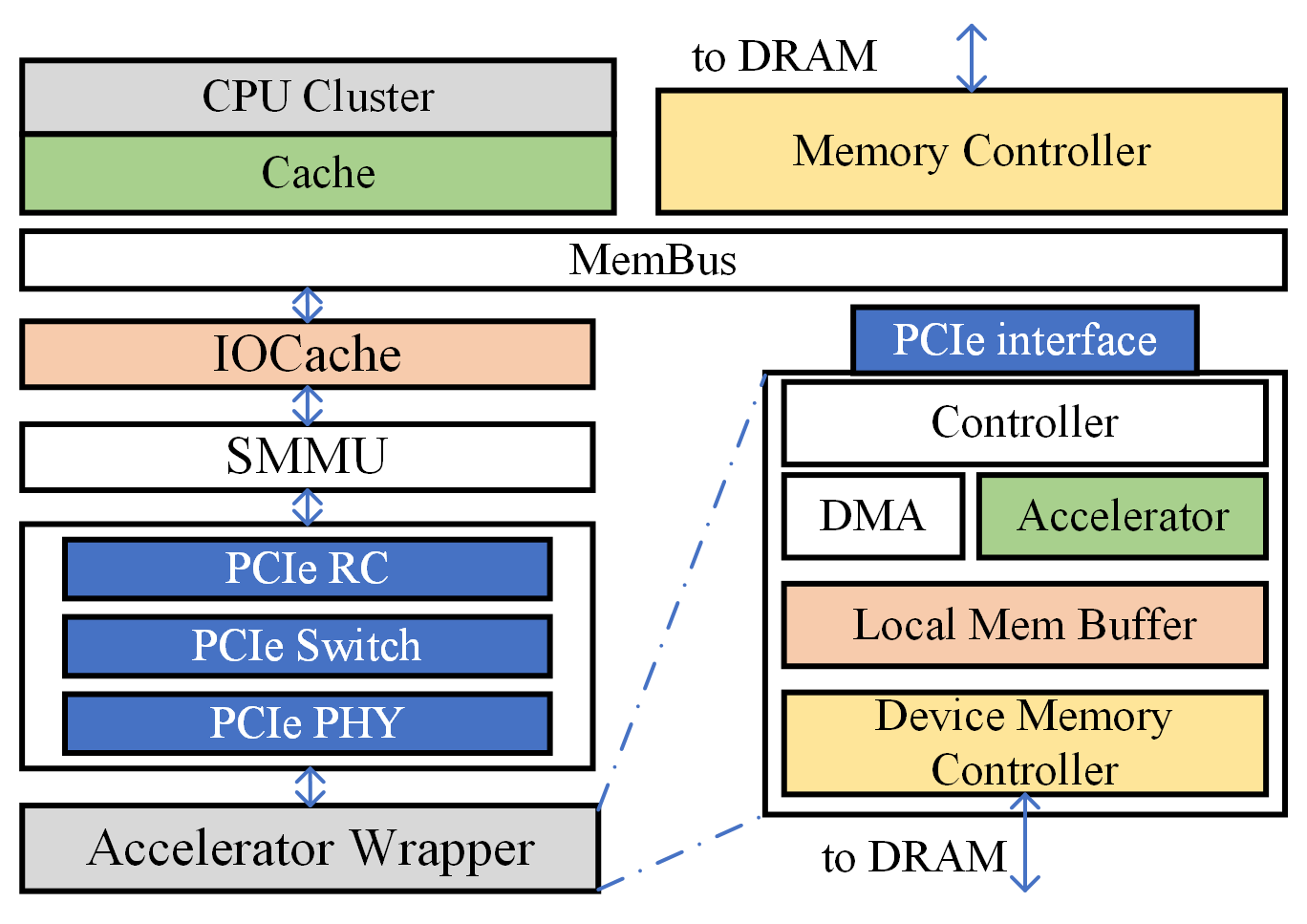}
  \caption{Design Framework Architecture}
  \label{fig:designFrame}
  \vspace{-0.5cm}
\end{figure}

Figure \ref{fig:designFrame} illustrates the architecture of our system. Modern computer systems incorporate not only CPU clusters but also complex interconnects for efficient data movement, including PCIe interfaces~\cite{b14}, Direct Memory Access (DMA), System Memory Management Units (SMMUs), etc. To fully exploit the performance of these features, we designed an accelerator wrapper optimized for data movement. The system is organized into two primary sections: the CPU cluster with its cache and the memory system interfaced with the accelerator through PCIe components. On the left side, we have the CPU cluster, which comprises one or more CPUs and their associated caches. This cluster is connected to the system's main memory (DRAM) through a memory controller. The connection to the memory controller is facilitated by a memory bus, which indicates the path for data transfers between the CPU and the memory. By attaching the accelerator to the computer system, we convert point-to-point data movements to pipelined data movements, where each pipeline stage contains suitable data buffers to stream the entire pipeline, as explained in Section \ref{sec:acceopt_mm}.

In addition, a kernel driver was developed to initiate accelerator operations, supporting full-system simulation. We provide two modes to explore the design space: Direct Memory (DM) access and Direct Cache (DC) access. In DM mode, requests are sent to the memory controller via the PCIe component with large granularity. In contrast, for DC, requests utilize finer granularity (64B) and are forwarded to the last-level cache before proceeding to the next memory level. The performance of these modes is analyzed in Section~\ref{sec:acceopt_mm}. By implementing the modified hardware system and the kernel driver, we integrate the accelerator as a loosely-coupled device.

\subsection{Accelerator-Optimized Matrix Multiplication Method}
\label{sec:acceopt_mm}
\begin{figure}[h]
\vspace{-0.2cm}
  \centering
  \vspace{-0.2cm}
  \includegraphics[width=\linewidth]{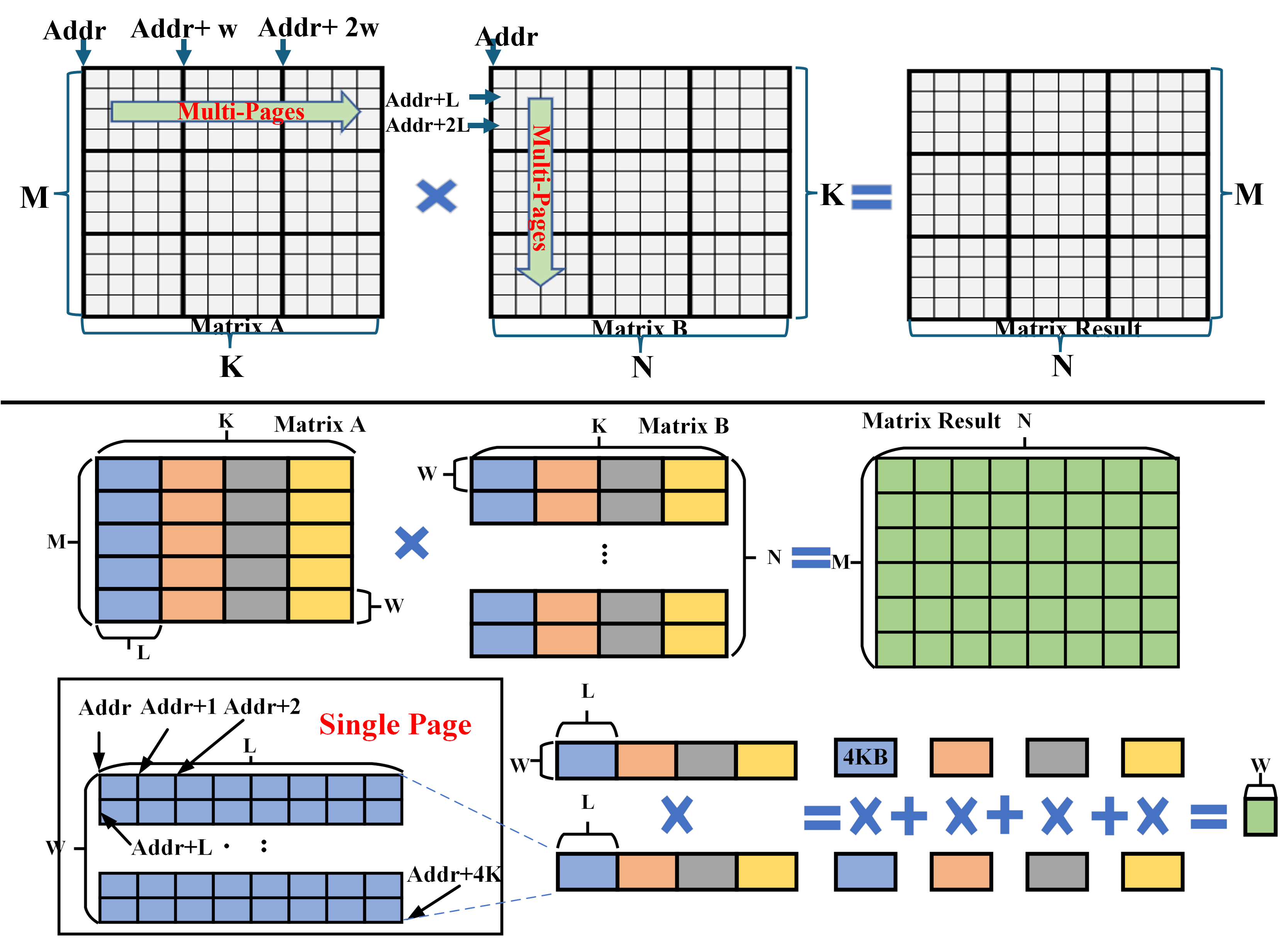}
  \caption{Conventional Matrix blocking and multiplication (top) and Our method for Matrix multiplication (bottom)}
  \label{fig:matrix_mul_combined}
  \vspace{-0.25cm}
\end{figure}

Traditional matrix multiplication methods used in existing SA designs introduce inefficiencies in data movement and memory access. As shown in Figure~\ref{fig:matrix_mul_combined} (top), Matrix~$A$ is typically read row-by-row, and Matrix~$B$ is read column by column. This approach conflicts with how modern systems manage memory, particularly for SAs. Although Matrix~$A$'s rows are stored contiguously, accessing blocks suitable for systolic arrays leads to non-contiguous memory accesses, causing inefficiencies due to fragmentation. For Matrix~$B$, columns are stored non-contiguously across memory pages, requiring multiple address translations and increasing overhead. Standard interconnects such as PCIe cannot fetch data from disparate pages simultaneously, which further exacerbates the problem.

To address these limitations, we propose a novel approach to data structure and memory management, illustrated in Figure~\ref{fig:matrix_mul_combined} (bottom). Our design splits the entire matrix into rectangular blocks that fit onto a single memory page (4KB). This block-based organization ensures that all data required for SA operations can be fetched with a single-memory access request. For Matrix$A$, the blocks are aligned with the SA's input dimensions, preserving address continuity. For Matrix~$B$, the blocks are restructured horizontally to align with the memory pages, resolving the non-contiguous address issues of column reads. This minimizes address translations and leverages PCIe capabilities for efficient data transfer. By fetching and processing a single block at a time, we reduce system overhead, eliminate repeated address translations, and accelerate throughput. This approach also optimizes pipeline utilization in SAs, allowing data to flow seamlessly without memory access delays.

Algorithm~\ref{alg:block_matrix_multiplication} outlines the block-based matrix multiplication process. The algorithm initializes the block dimensions and the result matrix~$C$, then iterates over the blocks of $A$ and $B$, multiplying and accumulating the results using the \textsc{MultiAcc} function. By implementing block-based multiplication, we convert conventional matrix multiplication into data flows of blocks. Each accelerator channel retrieves data from memory in a time-multiplexed manner, sharing the same PCIe link, switch, and root complex. Multiplexing channels maximizes system utilization, fully exploiting our novel data structure.

\begin{algorithm}
\caption{Optimized Block Matrix Multiplication}
\label{alg:block_matrix_multiplication}
\begin{algorithmic}[1]
    \STATE \textbf{function} \textsc{BlockMatrixMultiply}($A, B, M, N, K$)
    \STATE $Res \gets$ InitializeMatrix($M, N$)
    \STATE Divide $A$ and $B$ into blocks of size $W \times L$
    \FOR{$i = 0$ to $M / W - 1$}
        \FOR{$j = 0$ to $N / W - 1$}
            \STATE $Res_{block} \gets$ ZeroMatrix($W, W$)
            \FOR{$k = 0$ to $K / L - 1$}
                \STATE $A_{block} \gets$ GetBlock($A, i, k$)
                \STATE $B_{block} \gets$ GetBlock($B, j, k$)
                \STATE $Res_{block} \gets$ MultiAcc($A_{block}, B_{block}, Res_{block}$)
            \ENDFOR
            \STATE SetBlock($Res, i, j, Res_{block}$)
        \ENDFOR
    \ENDFOR
    \STATE \textbf{return} $Res$
\end{algorithmic}
\vspace{-0.1cm}
\end{algorithm}
\vspace{-0.25cm}

\begin{figure}[h]
\vspace{-0.4cm}
  \centering
  \includegraphics[width=\linewidth]{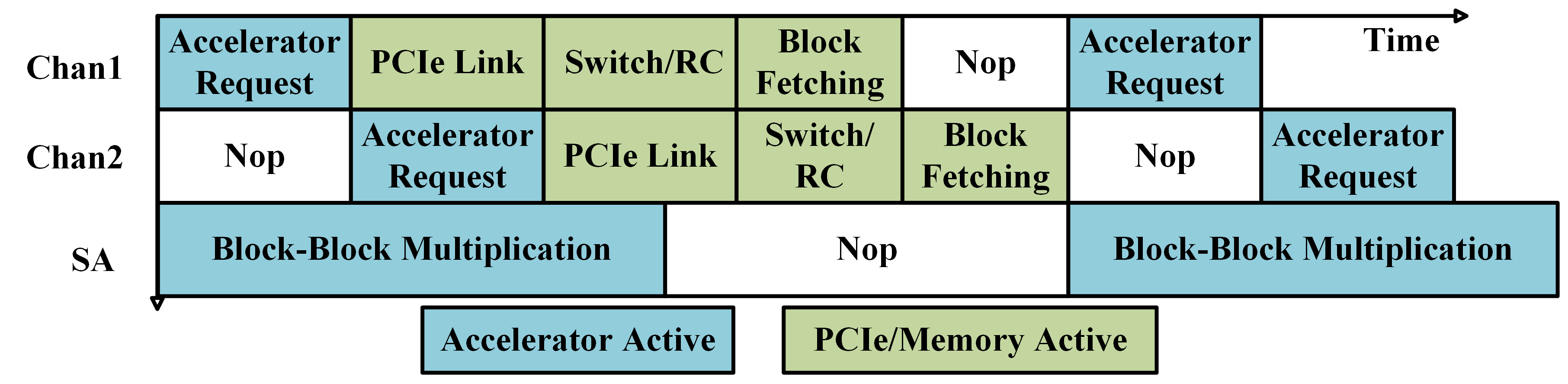}
  \vspace{-0.25cm}
  \caption{Matrix multiplication pipeline}
  \label{fig:matrix_pipeline}
\end{figure}

By integrating these methods, we convert conventional matrix multiplication into a data flow-based approach with page-sized blocks as the minimal unit, which frees the CPU from intensive instruction processing and data handling (see Figure~\ref{fig:matrix_pipeline}). This ensures continuous data feeding into the accelerator, maximizing throughput, and minimizing latency.
\textbf{Key Advancements:} \begin{itemize} \item \textbf{Block Segmentation:} Dividing matrices into accelerator-friendly blocks utilizes the maximum payload of interconnect systems and aligns with systolic array dimensions. \item \textbf{Memory Alignment:} Storing blocks to align with memory pages improves spatial locality and optimizes memory access patterns. \item \textbf{Horizontal Splitting:} Splitting matrices horizontally, unlike traditional vertical splitting, enhances data retrieval by aligning with memory architecture. \end{itemize}

By effectively addressing the limitations of conventional matrix multiplication methods, our approach demonstrates significant performance improvements, which makes it well-suited for high-throughput computing environments.

\section{Experimental results}

\subsection{Experimental setup}
Table~\ref{tab:system_configuration} describes the configuration of our simulated system, which consists on an ARM architecture equipped with a DDR3 memory and a PCIe interconnect.

\begin{table}[h]
\vspace{-0.3cm}
\centering
\begin{tabular}{lc}
\hline
\textbf{Component} & \textbf{Specification}\\
\hline
CPU & ARM, 1 GHz \\
Data Cache & 64 kB\\
Instruction Cache & 32 kB\\
Last Level Cache & 2 MB \\
Memory & DDR3\_1600\_8x8, 4 GB \\
PCIe & Version 6.0, 64 Gb/s, 16 Lanes \\
\hline
\end{tabular}
\caption{System Configuration}
\label{tab:system_configuration}
\vspace{-0.5cm}
\end{table}

The performance of our proposed accelerator is evaluated using two workloads: general matrix multiplication (GEMM) and transformers. 
We implement GEMM with matrix sizes ranging from \(64 \times 64\) to \(2048 \times 2048\) and compare the performance impacts of DC and DM.
For transformer models, we select BERT~\cite{b15} and Vision Transformer (ViT)~\cite{b16}. BERT is a deeply bidirectional transformer model for NLP pre-training implemented in three variants: medium, base, and large. ViT is a deep learning model architecture that applies transformers initially designed for NLP to image analysis.

\label{sec:ExperiResult}
To evaluate the performance of the algorithm and matrix, we implement the aforementioned workload on the system.
\subsection{Hardware result}
To support multiple data types, we designed dedicated hardware for the Multiply-Accumulate (MAC) units that compose the systolic array. The MAC units are arranged in a 16x16 array, which are synthesized using the TSMC 28nm HPC library. The following table summarizes the synthesis results for each data type. In addition, we analyze the power consumption of each MAC unit to assess the power impact on different data types. Table~\ref{tab:MAC_PPA} presents the area, frequency and power consumption for each type of data.

\begin{table}[ht]
\vspace{-0.2cm}
\centering
\begin{tabular}{cccccc}
\hline
\textbf{Metric}         & \textbf{INT32}        & \textbf{INT16}        & \textbf{INT8}         & \textbf{FP32}        & \textbf{FP16}    \\ \hline
Frequency      & 1GHz         & 1GHz         & 1GHz         & 600MHz      & 600MHz      \\
Area ($\text{mm}^2$) & 0.611      & 0.193      & 0.054       & 0.694        & 0.199     \\
Power (mW)    & 585.20    & 409.74  & 353.64            & 320.32          & 245.661          \\ \hline
\end{tabular}
\caption{PPA Results for Different Data Types}
\label{tab:MAC_PPA}
\vspace{-0.7cm}
\end{table}

\subsection{GEMM performance results}
To benchmark our work against existing methods, we use the performance of a single-threaded CPU that executes a GEMM implementation based on loops as a baseline. For a fair performance comparison, we also evaluate against a multi-threaded CPU implementation using OpenMP~\cite{b17} using 256 cores. Additionally, we analyze the performance improvements introduced by the Systolic Array Accelerators using Direct Cache (DC) and Direct Memory (DM) access methods.

\textbf{Multi-threading with OpenMP (OMP)} utilizes the parallel processing capabilities of modern multi-core processors. By distributing matrix multiplication tasks across multiple threads, OpenMP significantly reduces computation time. The results were obtained on a Cavium ThunderX2 server with up to 256 logical cores.

\textbf{ARM Neon} is a Single Instruction Multiple Data (SIMD) co-processor. Our code can be compiled with vectorization using the GCC compiler and launched directly into ARM Neon.


\textbf{Direct Cache (DC) Access} leverages the processor's last-level cache (LLC) to store data close to compute units, reducing the latency associated with retrieving data from main memory. 

\textbf{Direct Memory (DM) Access} bypasses the cache and directly transfers data between the main memory and the accelerator, alleviating cache pressure. DM uses a larger, adjustable burst length to retrieve data from the memory controller, making it effective for larger matrices with high cache pressure. 

\subsubsection{Performance improvement with varying matrix size}
\begin{figure}[h]
  \vspace{-0.4cm}
  \centering
  \includegraphics[width=\linewidth]{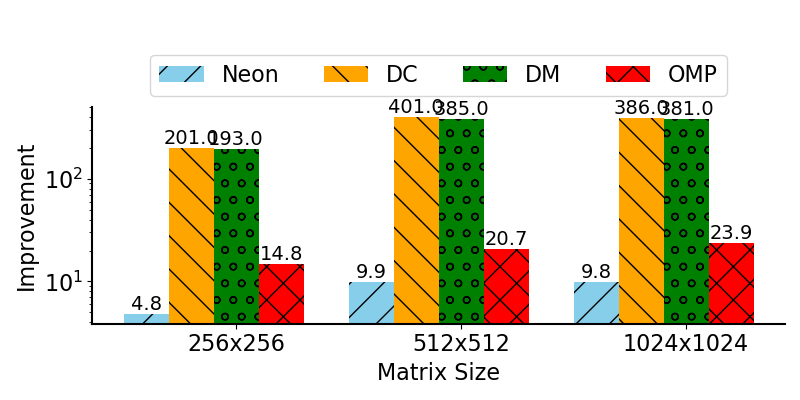}
  \vspace{-0.8cm}
  \caption{Matrix Multiplication Performance Comparison}
  \label{fig:matrix_performance}
\end{figure}

We implement GEMM with matrix sizes ranging from 256 to 1024, with Int8 precision. The results, summarized in Figure~\ref{fig:matrix_performance}, indicate significant performance improvements. GEMM utilizing DC outperforms the baseline, showing up to a 400x speed increase as matrix sizes increase. This superior performance is attributed to the higher bandwidth afforded. Following closely, DM achieves up to a 385x speed-up. Implementing the customized data structure significantly surpasses the baseline, demonstrating the impact of a well-suited data structure and algorithm. As expected, the for-loop-based algorithm exhibits the weakest performance, even with multi-threading.

\subsubsection{Performance improvement with varying date type}
To evaluate the performance of GEMM in different precisions, we conducted experiments using a variety of data types. The single-threaded CPU implementation serves as the baseline, while GEMM is executed on a multi-threaded CPU, ARM Neon, and the accelerator with both DC and DM access methods. Note that many ARM-based CPUs do not natively support float16 in Neon. As a result, the float16 implementation on the CPU requires additional data conversion, leading to increased cycle counts. Furthermore, the current Gem5 version does not support float16 on Neon, so we utilized float NEON to implement float16. In this experiment, a matrix with dimensions of 512 was used as a workload.

\begin{figure}[h]
  \vspace{-0.4cm}
  \centering
  \includegraphics[width=\linewidth]{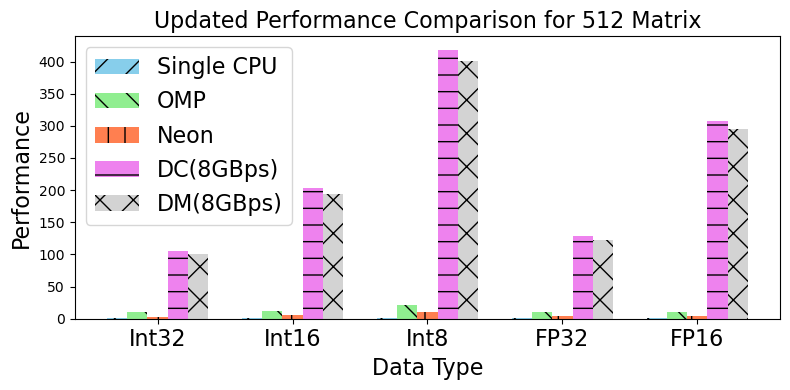}
  \vspace{-0.4cm}
  \caption{Matrix Multiplication Performance Comparison}
  \label{fig:perf_datatype}
\end{figure}

Figure~\ref{fig:perf_datatype} illustrates the performance comparison for various data types. The results reveal that fp16 delivers the most significant performance gains when executed on the accelerator. In the case of Neon, performance improves as the data-type size decreases, particularly for integer data types.

These findings underscore the effectiveness of hardware accelerators and parallel processing in handling computationally intensive tasks such as GEMM.
The subsequent sections explore latency and throughput considerations in more detail.

\subsection{Transformer performance results}

We implemented three different BERT and ViT configurations on the accelerator, focusing on DC, as it gives the best results for GEMM. The accelerator handles matrix multiplication, while the CPU manages other layers such as softmax, layer normalization, and transpose. We compare the performance of \emph{full transformer execution} using the same \textbf{Baseline} than in Section~\ref{sec:ExperiResult}. We compare also against \textbf{Neon} and \textbf{OMP} and we add two extra scenarios of state-of-the-art works using both loosely-coupled and tightly-coupled accelerators:
\begin{itemize} 
    \item \textbf{SMAUG~\cite{b18}} - A framework focusing on end-to-end simulation of DNN workloads, integrating custom hardware accelerators easily into the simulation environment in a loosely-coupled way. SMAUG has been shown to enhance performance and energy efficiency, achieving a speedup of up to 1.8x-5x over baseline systems.
    \item \textbf{Tightly-coupled SA~\cite{b2}} - These systolic arrays (TiC-SATs) include tightly-coupled, small-scale architectures with dedicated ISA extensions that significantly speed up execution. They employ software optimizations to maximize data reuse and reduce cache miss rates across cache hierarchies. 
    \item \textbf{MatrixFlow} - Our DC configuration of section~\ref{sec:ExperiResult}.
\end{itemize}

\begin{table}[htbp]
  \vspace{-0.2cm}
  \centering
  \resizebox{\columnwidth}{!}{
    \begin{tabular}{lcccccc}
      \toprule
      \textbf{Configuration} & \multicolumn{3}{c}{\textbf{BERT Models}} & \multicolumn{3}{c}{\textbf{ViT Models}} \\
      \cmidrule(lr){2-4} \cmidrule(lr){5-7}
      & \textbf{Medium} & \textbf{Base} & \textbf{Large} & \textbf{Base} & \textbf{Large} & \textbf{Huge} \\
      \midrule
      Single-thread   & 1    & 1    & 1    & 1    & 1    & 1 \\
      Multi-threaded  & 23.7 & 24.3 & 25.6 & 23.7 & 24.3 & 25.6 \\
      SMAUG~\cite{b2}           & 88   & -    & -    & -    & -    & - \\
      TIC-SAT~\cite{b2}         & 58.3 & 69.3 & 89.5 & 69.4 & 82.5 & 82.7 \\
      MatrixFlow      & 453.9& 633.7& 698.2& 327.9& 392.0& 427.6\\
      \bottomrule
    \end{tabular}
  }
  \caption{Transformer Performance Comparison}
  \label{tab:transperformance}
  \vspace{-0.5cm}
\end{table}

Table~\ref{tab:transperformance} presents the results of our experiments. We see how the performance of OMP is moderate when compared with other architectures and stagnates for large models, not scaling properly.

We also compare our results with the SMAUG platform. Both our implementation and SMAUG extend the gem5 simulator, enabling a fair comparison. The primary difference between our design and SMAUG is the targeted precision: SMAUG uses Float16, whereas our design targets Int32 given that we use a SA accelerator. SMAUG achieves a speed-up of up to 88x relative to the baseline~\cite{b18}.

The tightly-coupled accelerator TiC-SAT is also implemented as a gem5 extension and represents the closest work to ours. TiC-SAT incorporates an SA accelerator as a functional unit in the execution stage of the CPU pipeline. Customized instructions supporting the SA are implemented on the ARM ISA. TiC-SAT uses Int8 data type and shows a considerable performance improvement that scales well for large models. 
Our design not significantly outperforms the baseline, achieving a speed-up of up to 441.39x, but outperforms also all other methods, both loosely-coupled and tightly coupled, and scales well with increasing workload sizes. In the following section, we dig into the reasons behind these improvements.


\subsection{MatrixFlow Performance Analysis}


\subsubsection{Transformer Runtime Analysis}
\begin{figure}[h]
  \vspace{-0.3cm}
  \centering
  \includegraphics[width=\linewidth]{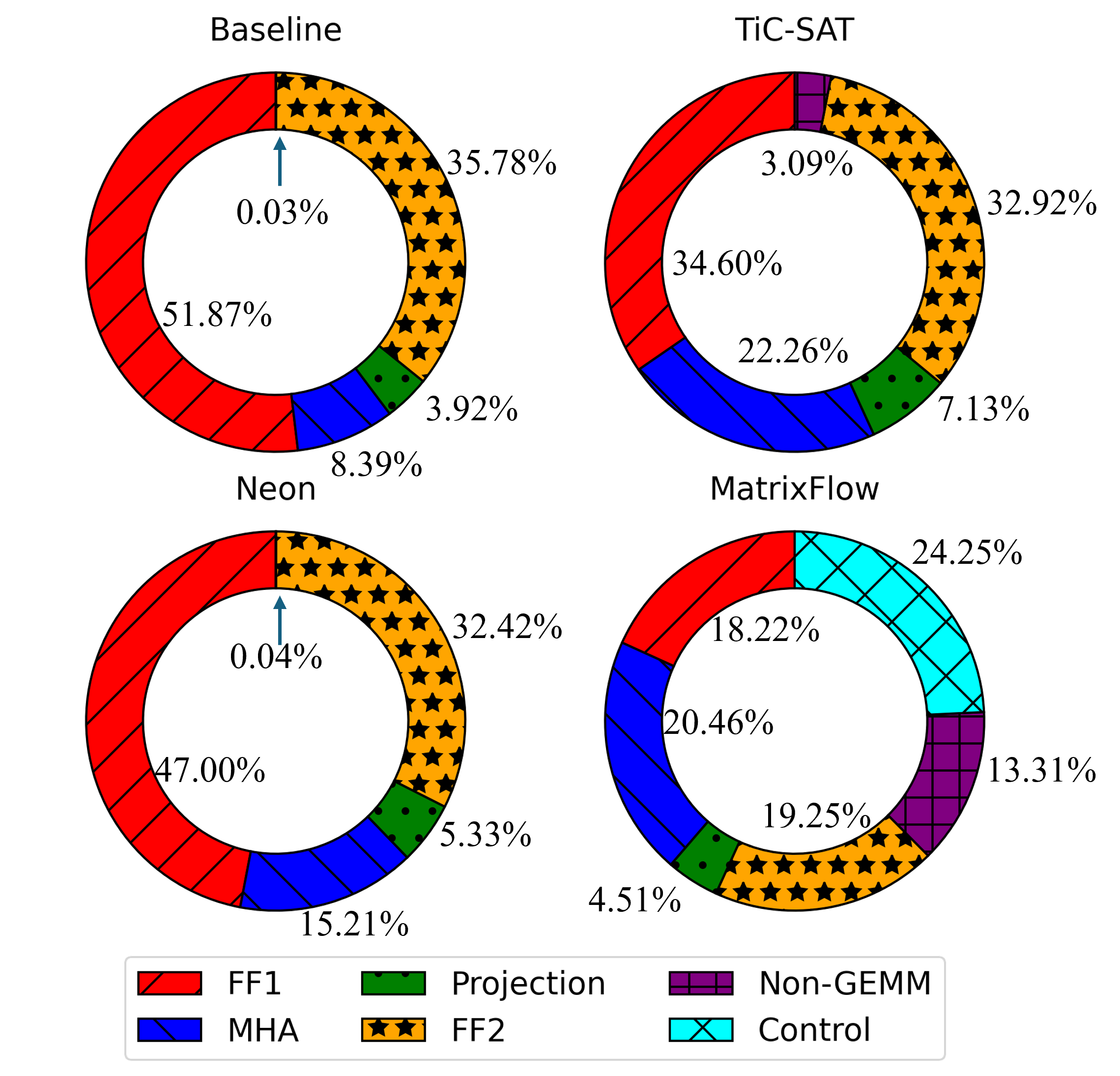}
  \label{fig:runtimeAna}
  \vspace{-0.4cm}
  \caption{Runtime Analysis}
\end{figure}

Figure~\ref{fig:runtimeAna} illustrates the results of different scenarios. The \textbf{Baseline} result represents the pure single-core CPU system without any optimization, which serves as our baseline. The \textbf{Neon} demonstrates the performance of the CPU with a customized data structure. In the baseline, GEMM operations account for most of the performance overhead, reaching 99\%. Within it, the fully-connected forward network is the most significant contributor, accounting for more than 87.7\%. Neon shows a reduced overhead for GEMM operations. For accelerated systems (TiC-SAT and MatrixFlow), the results differ significantly. As matrix multiplications are offloaded to the accelerator, its performance greatly improves compared to the other scenarios. For MatrixFlow, the enhanced data transmission  via DMA and PCIe led to an increase in the time spent on GEMM operations. In particular, non-GEMM computation still contributes considerably to the overall performance overhead, reaching 13.32\%. Moreover, due to the loosely coupled architecture, the control overhead, mainly from command reaching from CPU and descriptor fetching from memory to support DMA, is considered as well, which contributes to 24.25\%. It is worth noting that Multi-head attention (MHA) and Feed-Forward layer are still the main source of performance overhead. Compared to \textbf{TiC-SAT}, the overhead percentage of Feed-Forward layer 1 (FF1) and Feed-Forward layer 2 (FF2) decreased significantly, showcasing the ability to offload large matrix multiplication into the loosely-coupled accelerator.

\subsubsection{Impact of PCIe Speed}
A key element responsible for the superior performance of MatrixFlow is how it connects to the system. To evaluate the impact of PCIe configurations on system performance, a matrix multiplication workload was tested under various PCIe conditions: \textbf{16 lanes-64 Gbps}, \textbf{4 lanes-16 Gbps}, and \textbf{4 lanes-5 Gbps}.
Each configuration was assessed using DC to examine how bandwidth and lane count affect data throughput.

\begin{figure}[h]
  \vspace{-0.4cm}
  \centering
  \includegraphics[width=\linewidth]{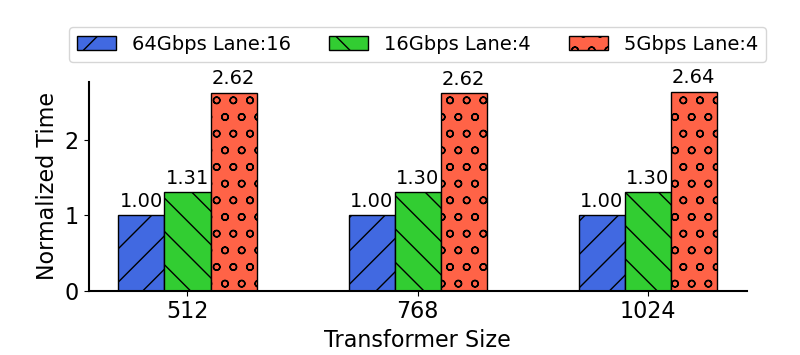}
  \vspace{-0.2cm}
  \caption{Performance Study with Different PCIe Bandwidth}
  \label{fig:pcie_study}
  \vspace{-0.4cm}
\end{figure}

As depicted in Figure~\ref{fig:pcie_study}, the performance varies significantly between configurations. The 16-lane, 64 Gbps setup outperforms the others, with results that are 130\% better compared to the 4-lane configuration at 5 Gbps. The 4 lanes in 16 Gbps setup show moderate performance, being 20\% worse than the highest but 50\% better than the lowest setup. These results underscore that higher lane counts and increased bandwidth substantially enhance performance, highlighting the critical role of PCIe specifications in optimizing data-intensive operations for AI accelerators.

\section{Conclusion}


Transformers are a key AI workload, and their computational demands necessitate the use of accelerators for efficient execution. In this paper, we have proposed MatrixFlow, a co-designed system-accelerator architecture that outperforms state-of-the-art designs (both loosely- and tightly-coupled) thanks to its hardware-software co-design approach. The new data structure and algorithm of MatrixFlow optimize the data movement with the help of standard hardware interconnects. The effective use of PCIe and DMA to fetch sizes of 4kB of data efficiently, minimizing data movement overhead, and outperforming other loosely coupled accelerators. Compared to tightly-coupled accelerators like TiC-SAT, our loosely-coupled design offers superior flexibility and scalability, resulting in performance enhancements that achieve 22x with respect to the many-core CPU baseline and over 4.5x and 8x improvements against the closest works in the state-of-the-art accelerator.
\bibliographystyle{ACM-Reference-Format}
\bibliography{sample-sigconf-authordraft}


\begin{thebibliography}{21}


\ifx \showCODEN    \undefined \def \showCODEN     #1{\unskip}     \fi
\ifx \showDOI      \undefined \def \showDOI       #1{#1}\fi
\ifx \showISBNx    \undefined \def \showISBNx     #1{\unskip}     \fi
\ifx \showISBNxiii \undefined \def \showISBNxiii  #1{\unskip}     \fi
\ifx \showISSN     \undefined \def \showISSN      #1{\unskip}     \fi
\ifx \showLCCN     \undefined \def \showLCCN      #1{\unskip}     \fi
\ifx \shownote     \undefined \def \shownote      #1{#1}          \fi
\ifx \showarticletitle \undefined \def \showarticletitle #1{#1}   \fi
\ifx \showURL      \undefined \def \showURL       {\relax}        \fi
\providecommand\bibfield[2]{#2}
\providecommand\bibinfo[2]{#2}
\providecommand\natexlab[1]{#1}
\providecommand\showeprint[2][]{arXiv:#2}

\bibitem[Alian et~al\mbox{.}(2018)]%
        {b14}
\bibfield{author}{\bibinfo{person}{M. Alian} {et~al\mbox{.}}} \bibinfo{year}{2018}\natexlab{}.
\newblock \showarticletitle{Simulating PCI-Express Interconnect for Future System Exploration}. In \bibinfo{booktitle}{\emph{Proceedings of IISWC 2018}}. \bibinfo{pages}{168--178}.
\newblock
\urldef\tempurl%
\url{https://doi.org/10.1109/IISWC.2018.8573496}
\showDOI{\tempurl}


\bibitem[Amirshahi et~al\mbox{.}(2023)]%
        {b2}
\bibfield{author}{\bibinfo{person}{A. Amirshahi} {et~al\mbox{.}}} \bibinfo{year}{2023}\natexlab{}.
\newblock \showarticletitle{TiC-SAT: Tightly-coupled Systolic Accelerator for Transformers}. In \bibinfo{booktitle}{\emph{Proceedings of ASP-DAC}}. \bibinfo{pages}{657--663}.
\newblock


\bibitem[Choi et~al\mbox{.}(2024)]%
        {b3}
\bibfield{author}{\bibinfo{person}{S. Choi} {et~al\mbox{.}}} \bibinfo{year}{2024}\natexlab{}.
\newblock \showarticletitle{SAVector: Vectored Systolic Arrays}.
\newblock \bibinfo{journal}{\emph{IEEE Access}}  \bibinfo{volume}{12} (\bibinfo{year}{2024}), \bibinfo{pages}{44446--44461}.
\newblock
\urldef\tempurl%
\url{https://doi.org/10.1109/ACCESS.2024.3380433}
\showDOI{\tempurl}


\bibitem[Dagum et~al\mbox{.}(1998)]%
        {b17}
\bibfield{author}{\bibinfo{person}{L. Dagum} {et~al\mbox{.}}} \bibinfo{year}{1998}\natexlab{}.
\newblock \showarticletitle{OpenMP: An Industry Standard API for Shared-memory Programming}.
\newblock \bibinfo{journal}{\emph{IEEE Computer Science \& Engineering}} \bibinfo{volume}{5}, \bibinfo{number}{1} (\bibinfo{year}{1998}), \bibinfo{pages}{46--55}.
\newblock
\urldef\tempurl%
\url{https://doi.org/10.1109/99.660313}
\showDOI{\tempurl}


\bibitem[Devlin et~al\mbox{.}(2019)]%
        {b15}
\bibfield{author}{\bibinfo{person}{J. Devlin} {et~al\mbox{.}}} \bibinfo{year}{2019}\natexlab{}.
\newblock \showarticletitle{BERT: Pre-training of Deep Bidirectional Transformers for Language Understanding}. In \bibinfo{booktitle}{\emph{Proceedings of NAACL-HLT 2019}}.
\newblock


\bibitem[Dosovitskiy et~al\mbox{.}(2021)]%
        {b16}
\bibfield{author}{\bibinfo{person}{Dosovitskiy} {et~al\mbox{.}}} \bibinfo{year}{2021}\natexlab{}.
\newblock \showarticletitle{An Image is Worth 16x16 Words: Transformers for Image Recognition at Scale}. In \bibinfo{booktitle}{\emph{Proceedings of ICLR 2021}}.
\newblock


\bibitem[Fornt et~al\mbox{.}(2023)]%
        {sysacc2}
\bibfield{author}{\bibinfo{person}{J. Fornt} {et~al\mbox{.}}} \bibinfo{year}{2023}\natexlab{}.
\newblock \showarticletitle{An Energy-Efficient GeMM-Based Convolution Accelerator With On-the-Fly im2col}.
\newblock \bibinfo{journal}{\emph{IEEE Transactions on Very Large Scale Integration (VLSI) Systems}} \bibinfo{volume}{31}, \bibinfo{number}{11} (\bibinfo{date}{nov} \bibinfo{year}{2023}), \bibinfo{pages}{1874--1878}.
\newblock


\bibitem[Genc et~al\mbox{.}(2021)]%
        {gemmini}
\bibfield{author}{\bibinfo{person}{Hasan Genc} {et~al\mbox{.}}} \bibinfo{year}{2021}\natexlab{}.
\newblock \showarticletitle{Gemmini: Enabling Systematic Deep-Learning Architecture Evaluation via Full-Stack Integration}. In \bibinfo{booktitle}{\emph{Proceedings of the 58th ACM/IEEE Design Automation Conference (DAC)}}. \bibinfo{publisher}{IEEE Press}, \bibinfo{pages}{769--774}.
\newblock
\urldef\tempurl%
\url{https://doi.org/10.1109/DAC18074.2021.9586216}
\showDOI{\tempurl}


\bibitem[Huang et~al\mbox{.}(2008)]%
        {gpu}
\bibfield{author}{\bibinfo{person}{Qihang Huang} {et~al\mbox{.}}} \bibinfo{year}{2008}\natexlab{}.
\newblock \showarticletitle{GPU as a General Purpose Computing Resource}. In \bibinfo{booktitle}{\emph{Proceedings of the PDCAT Conference}}.
\newblock


\bibitem[Ivanov et~al\mbox{.}(2020)]%
        {b9}
\bibfield{author}{\bibinfo{person}{A. Ivanov} {et~al\mbox{.}}} \bibinfo{year}{2020}\natexlab{}.
\newblock \showarticletitle{Data Movement Is All You Need: A Case Study of Transformer Networks}.
\newblock


\bibitem[Jouppi et~al\mbox{.}(2017)]%
        {b1}
\bibfield{author}{\bibinfo{person}{Norman~P. Jouppi} {et~al\mbox{.}}} \bibinfo{year}{2017}\natexlab{}.
\newblock \showarticletitle{In-Datacenter Performance Analysis of a Tensor Processing Unit}. In \bibinfo{booktitle}{\emph{Proceedings of ISCA '17}}. \bibinfo{publisher}{ACM}, \bibinfo{pages}{1--12}.
\newblock
\urldef\tempurl%
\url{https://doi.org/10.1145/3079856.3080246}
\showDOI{\tempurl}


\bibitem[Kung et~al\mbox{.}(1979)]%
        {b8}
\bibfield{author}{\bibinfo{person}{H.~T. Kung} {et~al\mbox{.}}} \bibinfo{year}{1979}\natexlab{}.
\newblock \showarticletitle{Systolic Arrays (for VLSI)}. In \bibinfo{booktitle}{\emph{Proceedings of the SIAM Conference on Sparse Matrix Processing}}. \bibinfo{publisher}{SIAM}.
\newblock


\bibitem[Miniskar et~al\mbox{.}(2023)]%
        {b10}
\bibfield{author}{\bibinfo{person}{N.~R. Miniskar} {et~al\mbox{.}}} \bibinfo{year}{2023}\natexlab{}.
\newblock \showarticletitle{Tiling Framework for Heterogeneous Computing of Matrix-based Tiled Algorithms}. In \bibinfo{booktitle}{\emph{Proceedings of ExHET '23}}. \bibinfo{publisher}{ACM}, \bibinfo{pages}{Article~1}.
\newblock
\urldef\tempurl%
\url{https://doi.org/10.1145/3587278.3595642}
\showDOI{\tempurl}


\bibitem[Okuda et~al\mbox{.}(2006)]%
        {b4}
\bibfield{author}{\bibinfo{person}{K. Okuda} {et~al\mbox{.}}} \bibinfo{year}{2006}\natexlab{}.
\newblock \showarticletitle{Reliable Systolic Computing Through Redundancy}. In \bibinfo{booktitle}{\emph{Proceedings of ACSAC 2006}} \emph{(\bibinfo{series}{LNCS}, Vol.~\bibinfo{volume}{4186})}. \bibinfo{publisher}{Springer}.
\newblock


\bibitem[Rock et~al\mbox{.}(2022)]%
        {b12}
\bibfield{author}{\bibinfo{person}{A. Rock} {et~al\mbox{.}}} \bibinfo{year}{2022}\natexlab{}.
\newblock \showarticletitle{INT8 Transformers for Inference Acceleration}. In \bibinfo{booktitle}{\emph{Proceedings of NeurIPS 2022}}.
\newblock


\bibitem[Shen et~al\mbox{.}(2019)]%
        {sysacc1}
\bibfield{author}{\bibinfo{person}{J. Shen} {et~al\mbox{.}}} \bibinfo{year}{2019}\natexlab{}.
\newblock \showarticletitle{A High-Performance Systolic Array Accelerator Dedicated for CNN}. In \bibinfo{booktitle}{\emph{Proceedings of the 2019 IEEE 19th International Conference on Communication Technology (ICCT)}}. \bibinfo{publisher}{IEEE}, \bibinfo{pages}{1200--1204}.
\newblock


\bibitem[Vaswani et~al\mbox{.}(2017)]%
        {b5}
\bibfield{author}{\bibinfo{person}{A. Vaswani} {et~al\mbox{.}}} \bibinfo{year}{2017}\natexlab{}.
\newblock \showarticletitle{Attention is All You Need}. In \bibinfo{booktitle}{\emph{Proceedings of NIPS '17}}. \bibinfo{publisher}{Curran Associates}, \bibinfo{pages}{6000--6010}.
\newblock


\bibitem[Wei et~al\mbox{.}(2017)]%
        {b7}
\bibfield{author}{\bibinfo{person}{X. Wei} {et~al\mbox{.}}} \bibinfo{year}{2017}\natexlab{}.
\newblock \showarticletitle{Automated Systolic Array Architecture Synthesis for High Throughput CNN Inference on FPGAs}. In \bibinfo{booktitle}{\emph{Proceedings of DAC 2017}}. \bibinfo{pages}{1--6}.
\newblock
\urldef\tempurl%
\url{https://doi.org/10.1145/3061639.3062207}
\showDOI{\tempurl}


\bibitem[Xi et~al\mbox{.}(2020)]%
        {b18}
\bibfield{author}{\bibinfo{person}{S. Xi} {et~al\mbox{.}}} \bibinfo{year}{2020}\natexlab{}.
\newblock \showarticletitle{SMAUG: End-to-End Full-Stack Simulation Infrastructure for Deep Learning Workloads}.
\newblock \bibinfo{journal}{\emph{ACM Transactions on Architecture and Code Optimization}} \bibinfo{volume}{17}, \bibinfo{number}{4} (\bibinfo{year}{2020}), \bibinfo{pages}{39}.
\newblock
\urldef\tempurl%
\url{https://doi.org/10.1145/3424669}
\showDOI{\tempurl}


\bibitem[Yu et~al\mbox{.}(2024)]%
        {vecpro}
\bibfield{author}{\bibinfo{person}{P. Yu} {et~al\mbox{.}}} \bibinfo{year}{2024}\natexlab{}.
\newblock \showarticletitle{An Energy Efficient Soft SIMD Microarchitecture and Its Application on Quantized CNNs}.
\newblock \bibinfo{journal}{\emph{IEEE Transactions on Very Large Scale Integration (VLSI) Systems}} \bibinfo{volume}{32}, \bibinfo{number}{6} (\bibinfo{year}{2024}), \bibinfo{pages}{1018--1031}.
\newblock


\bibitem[Yüzügüler et~al\mbox{.}(2023)]%
        {b6}
\bibfield{author}{\bibinfo{person}{A.~C. Yüzügüler} {et~al\mbox{.}}} \bibinfo{year}{2023}\natexlab{}.
\newblock \showarticletitle{Scale-out Systolic Arrays}.
\newblock \bibinfo{journal}{\emph{ACM Transactions on Architecture and Code Optimization}} \bibinfo{volume}{20}, \bibinfo{number}{2} (\bibinfo{year}{2023}), \bibinfo{pages}{27}.
\newblock
\urldef\tempurl%
\url{https://doi.org/10.1145/3572917}
\showDOI{\tempurl}


\end{thebibliography}

\end{document}